\documentclass[pra,twocolumn,superscriptaddress]{revtex4-2} 

\usepackage[a4paper,bindingoffset=0.2in,left=1cm,right=1.5cm,top=1in,bottom=1in,footskip=.25in]{geometry}
\usepackage[T1]{fontenc}
\usepackage[english]{babel}
\usepackage{siunitx}
\usepackage[utf8]{inputenc}
\usepackage{url}
\usepackage{dsfont}
\usepackage{bbm}
\usepackage{nicefrac}
\usepackage{setspace}
\usepackage[short]{optidef}
\usepackage{float}
\usepackage{comment}
\usepackage{dsfont}
\usepackage{amssymb}  
\usepackage{amsfonts}
\usepackage[colorlinks=true, citecolor=blue,urlcolor=blue,linkcolor=blue,filecolor=black]{hyperref}
\usepackage[colorinlistoftodos, color=green!40, prependcaption]{todonotes}
\usepackage{ulem}
\usepackage{amsthm}
\usepackage{mathtools}
\usepackage{physics}
\usepackage{graphicx}
\usepackage{adjustbox}
\usepackage{placeins}
\usepackage{lipsum}
\usepackage{csquotes}
\usepackage{booktabs}
\usepackage{url}
\newcommand{\beq}{\begin{equation}}
\newcommand{\eeq}{\end{equation}}
\renewcommand{\emph}{\textit}

\usepackage{soul}
\usepackage{graphicx}
\usepackage{dcolumn}
\usepackage{bm}

\usepackage[utf8]{inputenc}
\usepackage[T1]{fontenc}
\usepackage{mathptmx}

\begin{document}

\preprint{AIP/123-QED}

\title[]{Deploying an inter-European quantum network}



\author{Domenico Ribezzo}
\affiliation{Istituto Nazionale di Ottica del Consiglio Nazionale delle Ricerche (CNR-INO), 50125 Firenze, Italy}
\affiliation{Università degli Studi di Napoli Federico II, Napoli, Italy}

\author{Mujtaba Zahidy}
\affiliation{Center for Silicon Photonics for Optical Communication (SPOC), Department of Photonics Engineering, Technical University of Denmark, Kgs. Lyngby, Denmark}

\author{Ilaria Vagniluca}
\affiliation{QTI S.r.l.,  50125, Firenze, Italy}

\author{Nicola Biagi}
\affiliation{QTI S.r.l.,  50125, Firenze, Italy}

\author{Saverio Francesconi}
\affiliation{Istituto Nazionale di Ottica del Consiglio Nazionale delle Ricerche (CNR-INO), 50125 Firenze, Italy}
\affiliation{QTI S.r.l., 50125, Firenze, Italy}

\author{Tommaso Occhipinti}
\affiliation{QTI S.r.l.,  50125, Firenze, Italy}

\author{Leif K. Oxenløwe}
\affiliation{Center for Silicon Photonics for Optical Communication (SPOC), Department of Photonics Engineering, Technical University of Denmark, Kgs. Lyngby, Denmark}

\author{Martin Lon\v{c}ari\'c}
\affiliation{Centre of Excellence for Advanced Materials and Sensing Devices, Ru{\dj}er Bo\v{s}kovi\'c Institute, 10000 Zagreb, Croatia}

\author{Ivan Cviti\'c}
\affiliation{Department of Information and Communication Traffic, Faculty of Transport and Traffic Sciences, University of Zagreb, Croatia}

\author{Mario Stip\v{c}evi\'c}
\affiliation{Centre of Excellence for Advanced Materials and Sensing Devices, Ru{\dj}er Bo\v{s}kovi\'c Institute, 10000 Zagreb, Croatia}

\author{\v{Z}iga Pu\v{s}avec}
\affiliation{University of Ljubljana, Faculty of Mathematics and Physics, 1000 Ljubljana, Slovenia}

\author{Rainer Kaltenbaek}
\affiliation{University of Ljubljana, Faculty of Mathematics and Physics, 1000 Ljubljana, Slovenia}
\affiliation{IQOQI - Austrian Academy of Sciences, 1090 Vienna, Austria}

\author{Anton Ram\v{s}ak}
\affiliation{University of Ljubljana, Faculty of Mathematics and Physics, 1000 Ljubljana, Slovenia}

\author{Francesco Cesa}
\affiliation{Department of Physics, University of Trieste, Trieste, Italy}

\author{Giorgio Giorgetti}
\affiliation{ICT service area, University of Trieste, Trieste, Italy}

\author{Francesco Scazza}
\affiliation{Department of Physics, University of Trieste, Trieste, Italy}
\affiliation{Istituto Nazionale di Ottica del Consiglio Nazionale delle Ricerche (CNR-INO), 50125 Firenze, Italy}

\author{Angelo Bassi}
\affiliation{Department of Physics, University of Trieste, Trieste, Italy}

\author{Paolo De~Natale}
\affiliation{Istituto Nazionale di Ottica del Consiglio Nazionale delle Ricerche (CNR-INO), 50125 Firenze, Italy}

\author{Francesco Saverio Cataliotti}
\affiliation{Istituto Nazionale di Ottica del Consiglio Nazionale delle Ricerche (CNR-INO), 50125 Firenze, Italy}
\affiliation{Department of Physics, Università degli Studi di Firenze, Firenze, Italy}

\author{Massimo Inguscio}
\affiliation{Department of Engineering, Campus Bio-Medico University of Rome, 00128 Rome, Italy}
\affiliation{Istituto Nazionale di Ottica del Consiglio Nazionale delle Ricerche (CNR-INO), 50125 Firenze, Italy}
\affiliation{QTI S.r.l.,  50125, Firenze, Italy}

\author{Davide Bacco}
\email{dabac@fotonik.dtu.dk}
\affiliation{Center for Silicon Photonics for Optical Communication (SPOC), Department of Photonics Engineering, Technical University of Denmark, Kgs. Lyngby, Denmark}
\affiliation{QTI S.r.l.,  50125, Firenze, Italy}

\author{Alessandro Zavatta}
\email{alessandro.zavatta@ino.cnr.it}
\affiliation{Istituto Nazionale di Ottica del Consiglio Nazionale delle Ricerche (CNR-INO), 50125 Firenze, Italy}
\affiliation{QTI S.r.l.,  50125, Firenze, Italy}


\begin{abstract}
Around forty years have passed since the first pioneering works have introduced the  possibility of using quantum physics to strongly enhance communications safety. Nowadays Quantum Cryptography, and in particular, Quantum Key Distribution (QKD) exited the physics laboratories  to become commercial technologies that increasingly trigger the attention of States, military forces, banks, and private corporations. This work takes on the challenge of bringing QKD closer to a consumer technology: optical fibers deployed and used by telecommunication companies of different States have been used to realize a quantum network, the first-ever connecting three different countries. This pushes towards the necessary  coexistence of QKD and classical communications on the same infrastructure, which currently represents a main limit of this technology. Our network connects Trieste to Rijeka and  Ljubljana via a trusted node in Postojna; a key rate of over 3 kbps has been achieved in the shortest link, and a 7-hour long measurement has demonstrated the system stability and reliability. Finally, the network has been used for a public demonstration of QKD at the G20 Digital Ministers' Meeting in Trieste. The reported experimental results, together with the significant interest that one of the most important events of international politics has attracted, showcase the maturity of the QKD technology bundle, placing it in the spotlight for consumer applications in the near term. 
\end{abstract}
\maketitle

\section{Introduction}
The amount of internet traffic is strongly increasing every year, with 5.3 billion internet users expected by 2023 \cite{cisco_2}; in the same way, the number of breaches and total records exposed per breach continues to grow as well as the average cost of lost or stolen records \cite{cisco_2}. In this scenario, realizing  a worldwide quantum network that guarantees strong safety in communications is of utmost importance. Quantum Key Distribution (QKD), proposed by Bennett and Brassard in 1984, is a protocol that can provide unconditionally secure data communications enabled by the laws of quantum physics \cite{BB84,Scarani2009,Diamanti2016}. QKD is the most mature quantum-enabled technology, and multiple Countries have already implemented practical use-cases worldwide. For example,  optical fiber links \cite{chen_2021,dynes_2019,wengerowsky2019,joshi2020trusted}, satellites \cite{yin_2020}, or both, have been used to create a quantum network enhancing secure communications among different cities and two different Countries \cite{liao_2018}. It is worth noticing that fiber communications over  thousands of kilometers are possible only thanks to the trusted-node scenario (quantum states are measured and then subsequently re-encoded)~\cite{neumann2021experimentally,bacco2021toward,zhang2009megabits,chen2020sending,liu2021field}. In this way, it is possible to extend the maximum haul of a point-to-point link and to allow connection of multiple users \cite{VERGOOSSEN2020164}.
However, the long-term goal of a unified quantum network across the entire world is hampered by practical difficulties (i.e., different fiber infrastructures, different telecom operators, etc.) of interconnecting more countries through the already existing fibre infrastructure. In this context, one of the goals of the European Quantum Communication Infrastructure (EuroQCI) project \cite{euroqci_2} is to establish a European quantum network, able to overcome current limitations. 

In this work, we kicked-off the EuroQCI initiative by connecting Italy, Slovenia, and Croatia, three different European countries, over an in-fiber quantum network. A BB84 protocol using a time-bin encoding scheme and 1-decoy state method has been used for  the different links~\cite{bacco2019}.
The measured key rates in the two links Trieste-Postojna and Ljubljana-Postojna are over 2.0 kbps and 3.1 kbps, respectively, while the key rate in the high-loss link Trieste-Rijeka (25 dB) is 610 bps. The distributed quantum keys have been used to secure a virtual private network (VPN) among the users, which was employed for quantum-secured video-calls during the G20 event held in Trieste.

\section{Network architecture}
\begin{figure*}
    \centering
    \includegraphics[width=0.92\textwidth]{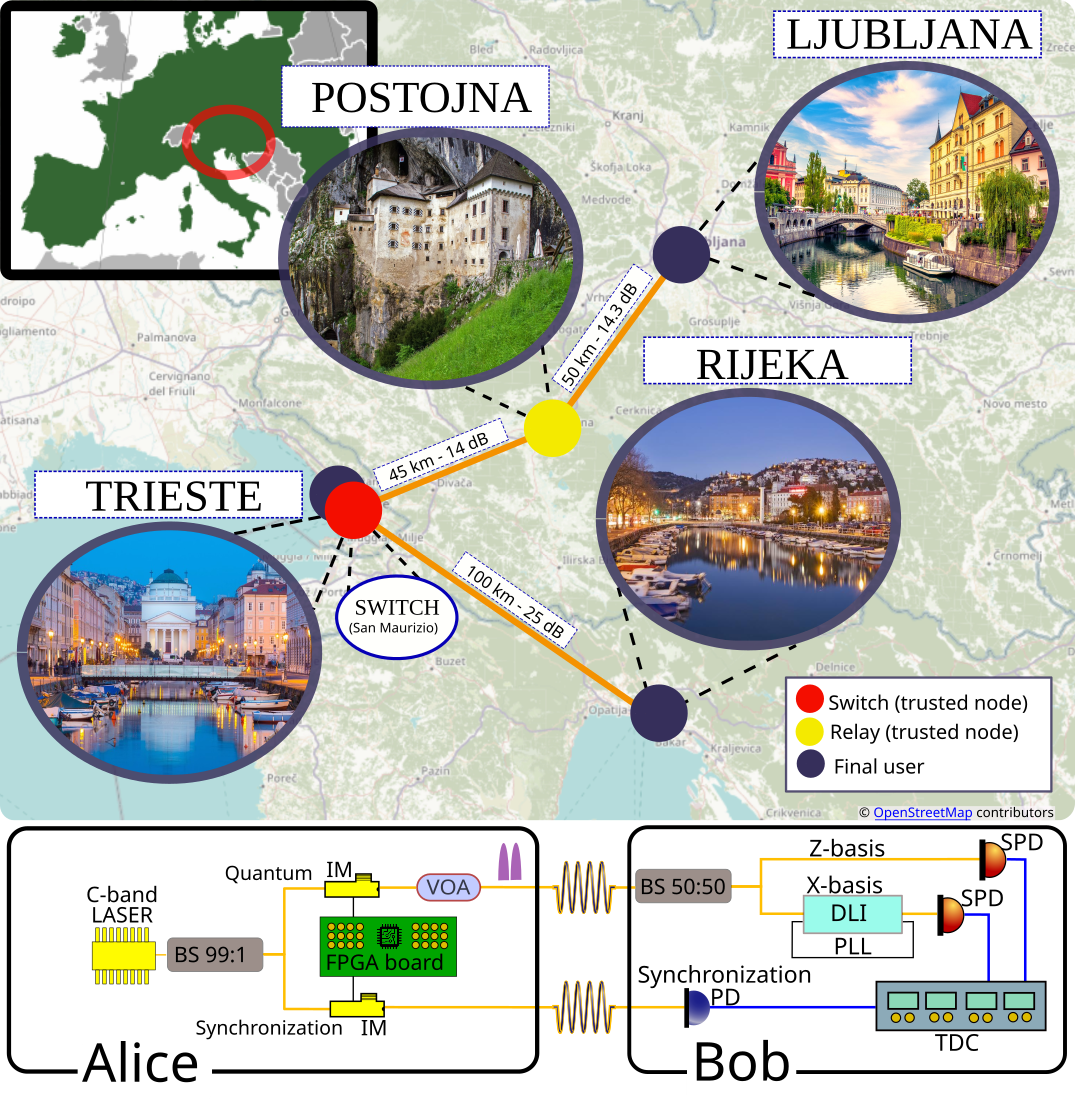}
    \caption{\textbf{Network and setup schemes.} \textbf{Top}: The network consists of three links and two trusted nodes. The two transmitters (Alices) are located in Trieste (Italy) and Ljubljana (Slovenia) while there are two receivers (Bobs) in Postojna (Slovenia) and one in Rijeka (Croatia). One transmitter is located in the Trieste Convention Center while a switch acting as a trusted node is located a few kilometers away, in the telecom center of Trieste San Maurizio; this solution renders the San Maurizio center the starting point of the quantum communication channels. The two receivers in Postojna behave as the second trusted node. \textbf{Bottom}: On Alice's side, the C-band laser is split by a 99:1 beam splitter (BS 99:1) and is sent to the intensity modulators (IM) controlled by an FPGA board; the 1\% output of the beam splitter goes toward the quantum part, while a variable optical attenuation (VOA) is added in order to reach the desired mean photon number per pulse. The 99\% BS output is used to generate a low-jitter synchronization signal with a frequency of 145 kHz. On Bob's side, a 50:50 BS is used for the basis choice; for the Z-basis the photons are directly sent to a single photon detector (SPD), while the photons directed to the X-basis detector pass previously through a delay-line interferometer (DLI), whose function is described in the text; in the Trieste-Postojna link, the interferometer is stabilized by a phase lock loop (PLL). The two SPDs, together with a fast photo-diode that reads the synchronization signal (sync PD), are connected to a time-to-digital converter (TDC) that registers the timestamps of events from which, after the post-processing stage, the key is extracted.}
    \label{fig:scheme}
\end{figure*}

\begin{figure}
    \centering
    
    \includegraphics[width=0.95\columnwidth]{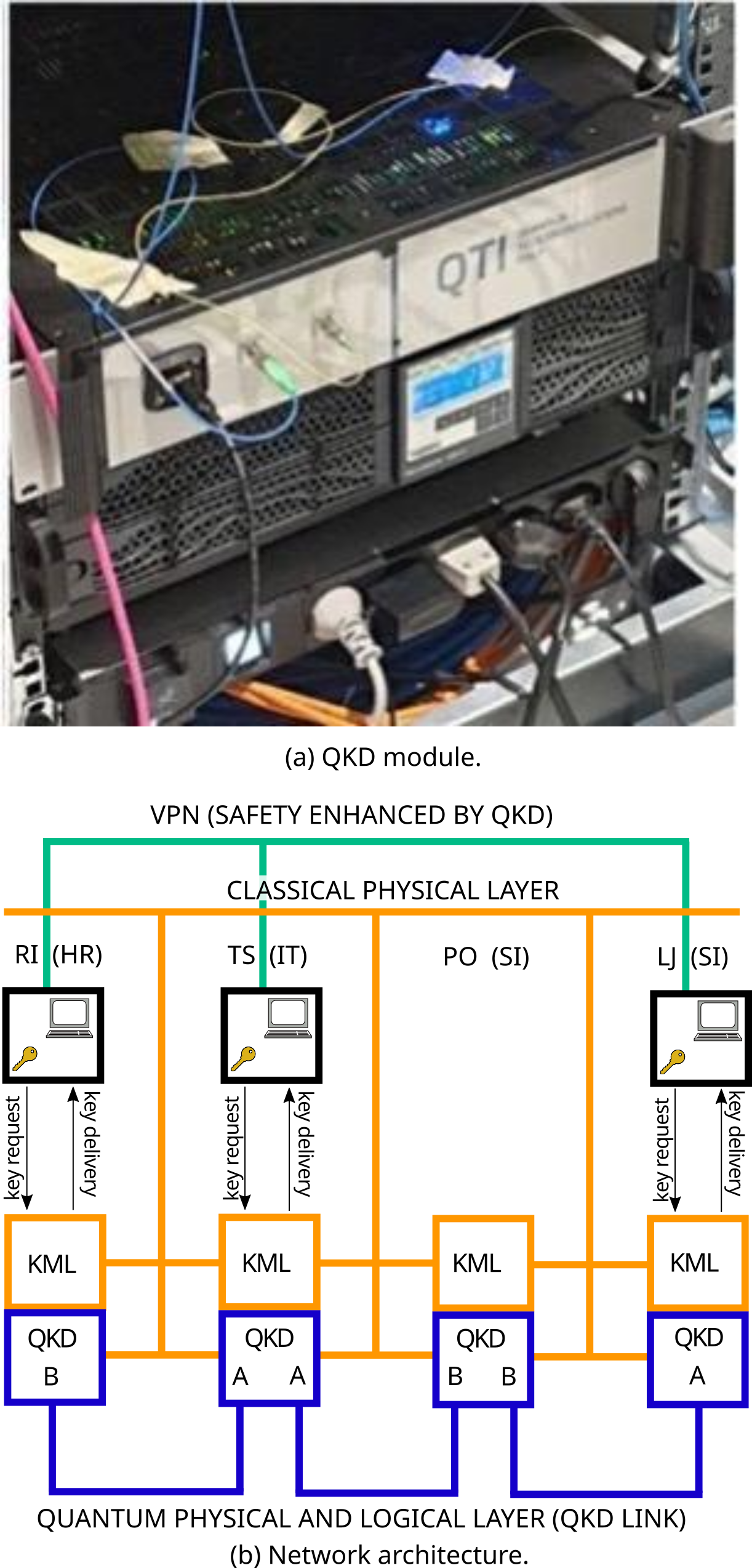}
    \caption{\textbf{a) QKD module.} QKD module mounted in a telecommunication cabinet. \textbf{b) Network architecture.} The network works over different levels.  The black squares, representing the computers in the meeting rooms, act as both the application layer and the classical logical layer; upon a request to launch the VPN connecting Trieste (TS), Rijeka (RI) and Ljubljana (LJ) nodes. The classical logical layer sends a request to the key management layer (KML), that checks if a ready-to-use key to secure the VPN is already stored. Upon passing this test the key management layer sends the key to the classical logical layer, otherwise, a request to generate a new key is sent to the quantum layer (blue box). The quantum layer is made by the proper optical setup (physical sublayer) and all the post-processing methods necessary to produce the final key (logical sublayer); the A (Alice) or B (Bob) letter in the quantum layer box says if the node is a transmitter or a receiver. When a key is ready, the quantum layer sends it to the key management layer that will deliver it to the first layer. The key management layer, together with the internet infrastructure, make up the classical physical layer.}
    \label{fig:layers}
\end{figure}
The implemented network, whose infrastructure is illustrated in Fig.~\ref{fig:scheme} and architecture is reported in Fig.~\ref{fig:layers}, is composed of two transmitters, also called \emph{Alice}, and three receivers, known as \emph{Bob}, connected by two optical fibers; one is used for the quantum signal and the other one is used for the service signal, i.e. synchronization, parameter estimation, etc. The first transmitter, located in Trieste Convention Center (TCC), sends the synchronization and the quantum signals encoding the key, towards the telecom center located in Trieste San Maurizio, three kilometers away from TCC. Here, the two signals are divided by two 50:50 beam splitters in order to route them towards two different nodes; this makes San Maurizio the place where the  quantum channels start. The two receiving nodes are both located outside the Italian borders, one in the Telekom Slovenije d.d. telecom center in Postojna (Slovenia) and the other one in the OIV telecom center located in Rijeka (Croatia). The Postojna node is not the final user of our network, since it acts as the second trusted node in order to reach the capital city,  Ljubljana, a connection  not possible  with a single direct link since the overall losses were too high (around 30 dB). More specifically, a second Alice was located at the Faculty of Mathematics and Physics of the University of Ljubljana: it worked in a way analogous to the Alice located in Trieste but served just one node. For each link, one dark fiber was used for the quantum channel and a second dark fiber for the synchronization. The communication for the upper layer protocols was established with a standard TCP/IP internet connection.
The measured attenuations of the quantum channels were about 14 dB for the links connecting Trieste to Postojna and Ljubljana to Postojna, and 25 dB for the Trieste-Rijeka link. The entire network has been set up in few days, from scratch, using already-deployed fibers from different providers. This quantum network was used to provide a QKD proof-of-principle demonstration at the G20 Digital Ministers' Meeting held in Trieste on August $5^{th}$, 2021. Two concerts by the Ljubljana and Rijeka conservatories orchestras were broadcasted to the G20 headquarter in Trieste via video-conference (openmeetings) established over a Virtual Private Network (VPN) reinforced  by a quantum key; likewise, the Trieste conservatory orchestra shared a concert with Ljubljana and Rijeka. 
The quantum network was based on fibers normally used for backup links and regular data traffic, and was hence only available for the amount of time necessary to configure the QKD setups and broadcast the three
concerts.

\section{QKD systems}
\subsection{Protocol}
The adopted protocol, implemented for all the links, is the three-states efficient BB84 with one decoy method \cite{boaron_2018,molotkov_nazin_1996,shi2000quantum,fung2006security,Rusca2018_SecProofSimpleBB84,Rusca2018_FiniteKeyAnalysis}. In this scheme, the key is distributed just using the states encoded in the $\mathbf{Z}$-basis, while the $\mathbf{X}$-basis states are utilized to estimate the amount of information leaked by an eavesdropper.
The mean numbers of photons per pulse contained in the signal and decoy states, generally referred to as $\mu_1$ and $\mu_2$, have been chosen in order to maximize the secure key rate, taking into account all the device characteristics. In the finite-key regime the key length $l$ is bound to \cite{boaron_2018}:
\small
\begin{equation}
    l\leq s_{Z,0}^l+s_{Z,1}^l(1-H_2(\phi_Z^u))-\lambda_{EC}-6\log_2\left(\frac{19}{\epsilon_{sec}}\right)-\log_2\left(\frac{2}{\epsilon _{corr}}\right),
    \label{EQ::}
\end{equation}
\normalsize
where $s_{Z,0}^l$ and $s_{Z,1}^l$ are the lower bounds for the vacuum and the single-photon events, respectively, $\phi_Z^u$ is the upper bound of the phase error rate, $\lambda_{EC}$ is the number of disclosed bits in the error correction stage, $H_2(x)=-x\log_2(x)-(1-x)\log_2(1-x)$ is the binary entropy and $\epsilon_{sec}$ and $\epsilon_{corr}$ are the secrecy and correctness parameters, defined as \cite{canale2014_2}:
\begin{equation*}
    \begin{split}
        P[S_A\neq S_B]  & < \epsilon_{corr}, \\
        \mathds{1}(S_A,S_B;Z,C) & < \epsilon_{sec},
    \end{split}
\end{equation*}
with $S_A$ and $S_B$ being the two sifted keys, $P[x]$ the probability of $x$, $\mathds{1}(\cdot)$ a generic information measure, $Z$ is the eavesdropped sequence owned by a potentially malicious part (generally referred to as Eve) and $C$ is a random variable representing the exchanged information. The second term denotes the probability $\epsilon_{sec}$ of having a stronger correlation between Alice's and Eve's strings than Alice's and Bob's one. These parameters have been arbitrarily set as $\epsilon_{corr}=10^{-12}$ and $\epsilon_{sec}=10^{-9}$. The phase error rate in the $\mathbf{Z}$-basis $\phi_Z$ can generally be estimated from the error rate in the $\mathbf{X}$-basis $\delta_X$; however, in the three-state BB84 protocol Alice sends only one quantum state in the $\mathbf{X}$-basis, so it cannot be directly measured but it needs to be estimated by the $\mathbf{X}$-basis error rate $QBER_X$ as reported in \cite{boaron2016detector}. $QBER_X$ is connected to the visibility $vis_X$ by $QBER_X=(1-vis_X)/2$.

\subsection{Alice's setup}
On Alice's side, a field-programmable gate array (FPGA) is programmed to generate digital signals that drive two intensity modulators, one to prepare the synchronization pulses at a rate of 145.358 kHz via carving a continuous wave (CW) laser, while the other one is producing the time-bin pulses, encoding the quantum states at a rate of around 595 MHz with 800 ps separation between the two bins. The sequence of quantum states is generated according to a pseudo-random binary sequence (PRBS) of length $l = 2^{12}-1$. The two CW laser beams are derived from the same C-band laser, previously divided by a beam-splitter. Alice's entire setup has been accurately engineered in order to fit in a 2U rack box. Alice makes the basis choice, encoding the qubit into the $\mathbf{Z}$ (computational) basis with a probability $P_{ZA}=0.9048$ or choosing the $\mathbf{X}$ (diagonal) basis with a probability $P_{XA}=1-P_{ZA}=0.0952$. These parameters have also been optimized according to the results of a simulation model, in order to maximize the final secret key rate given the channel loss and the detector efficiencies.
It should be noted that, in a real implementation of QKD, the quantum states are required to be phase randomized and the PRBS should be replaced with true random numbers \cite{Zahidy_2021_QRNG}.

\subsection{Bob's setup}
The qubits received by Bob are randomly measured in $\mathbf{Z}$ or $\mathbf{X}$-basis with equal probability through a 3 dB beam splitter. Concerning the $\mathbf{Z}$-basis, measuring the arrival time of the photon is sufficient for decoding the quantum state, while an unbalanced interferometric measurement with a delay line of 800 ps provides the relative phase information, performing the $\mathbf{X}$-basis measurement. Two types of interferometers have been used in Bob's setups: a free-space interferometer and a compact all-in-fiber one. The all-in-fiber interferometer is based on a Michelson interferometer with an additional 400 ps delay line in one arm and consists of two Faraday mirrors, a piezo-electric phase shifter, and an adjustable delay line to precisely match the length. The two arms are phase-stabilized with a phase-lock-loop (PLL) that drives the phase shifter according to the feedback provided by monitoring the phase fluctuation of a monitor laser (PLL laser). The PLL laser is mixed with the quantum signal via a dense wavelength division multiplexing device (DWDM) and is sent to the interferometer in a co-propagating way with respect to the quantum signal. A second DWDM separates the quantum signals from the PLL laser, where the latter is monitored with an avalanche photo-diode and provides feedback to the PLL.

The second interferometer \cite{florence_field_trial} is a free-space Mach-Zehnder interferometer, whose arms are again delayed 800 ps with respect to each other. It is equipped with a piezo-electric driven mirror to adjust the phase drifts, however, due to the high stability of the system in the periods of data acquisition, no active stabilization system has been put in place.

Measurements have been performed with single-photon detectors from ID quantique, MPD and Aurea able to provide up to 20\% efficiency, and dark counts in the range of 2500 Hz at 20 $\mu s$ hold-off time. Detection events and their arrival time have then been logged by Qutools QuTAG and Swabian Ultra time-to-digital converters, which both have  a temporal resolution of 1 ps and an RMS jitter lower than 10 ps.

\section{Results}
Table \ref{results} summarizes the important parameters and final secret key rates achieved in each link.
We began by characterizing the channel loss in each link and proceeded to choose the optimal mean photon numbers that maximize the final secret key rate. The mean photon number for signal ($\mu_1$) and decoy ($\mu_2$) state in Ljubljana-Postojna link has been set to $\mu_1=0.15$ and $\mu_2=0.06$, while for the source in Trieste, that served two nodes, the mean photon numbers per pulse have been chosen as the best compromise between the two channels. 


The average SKR per link obtained is around 2080 bps and 3130 bps for the Trieste-Postojna and the Ljubljana-Postojna link, respectively, while it was found around 610 bps for the longer and high-loss channel of the Trieste-Rijeka link. In all cases, we selected a block size such that the computational time necessary for the post-processing was not longer than the corresponding data acquisition time (\emph{block time}).


Adaptive temporal filters were utilized to compensate for the loss of signal, and for little temporal drift of the pulses due to thermal fluctuations in the fibers as well as to reduce the impact of background noise in measuring the visibility. The systems suffered from $\approx 2.5$ kcounts/s dark counts in each basis and the loss of the interferometer introduced by design. Due to the high level of attenuation experienced by the signal in Trieste-Rijeka channel and the low number of counts as a result of that, a 60 ps wide temporal filter was adopted to guarantee a QBER around $5\%$ in the security check basis, while the temporal filters for Trieste-Postojna and Ljubljana-Postojna were set to 100 ps and 200 ps, respectively. The difference is due to different losses in the measurement apparatus, indeed, the Michelson interferometer introduces more losses than the free-space Mach-Zehnder one, thus causing a lower signal-to-noise ratio in the TS-PO channel.

\begin{figure}[htbp]
    \centering
    \includegraphics[width=0.98\columnwidth]{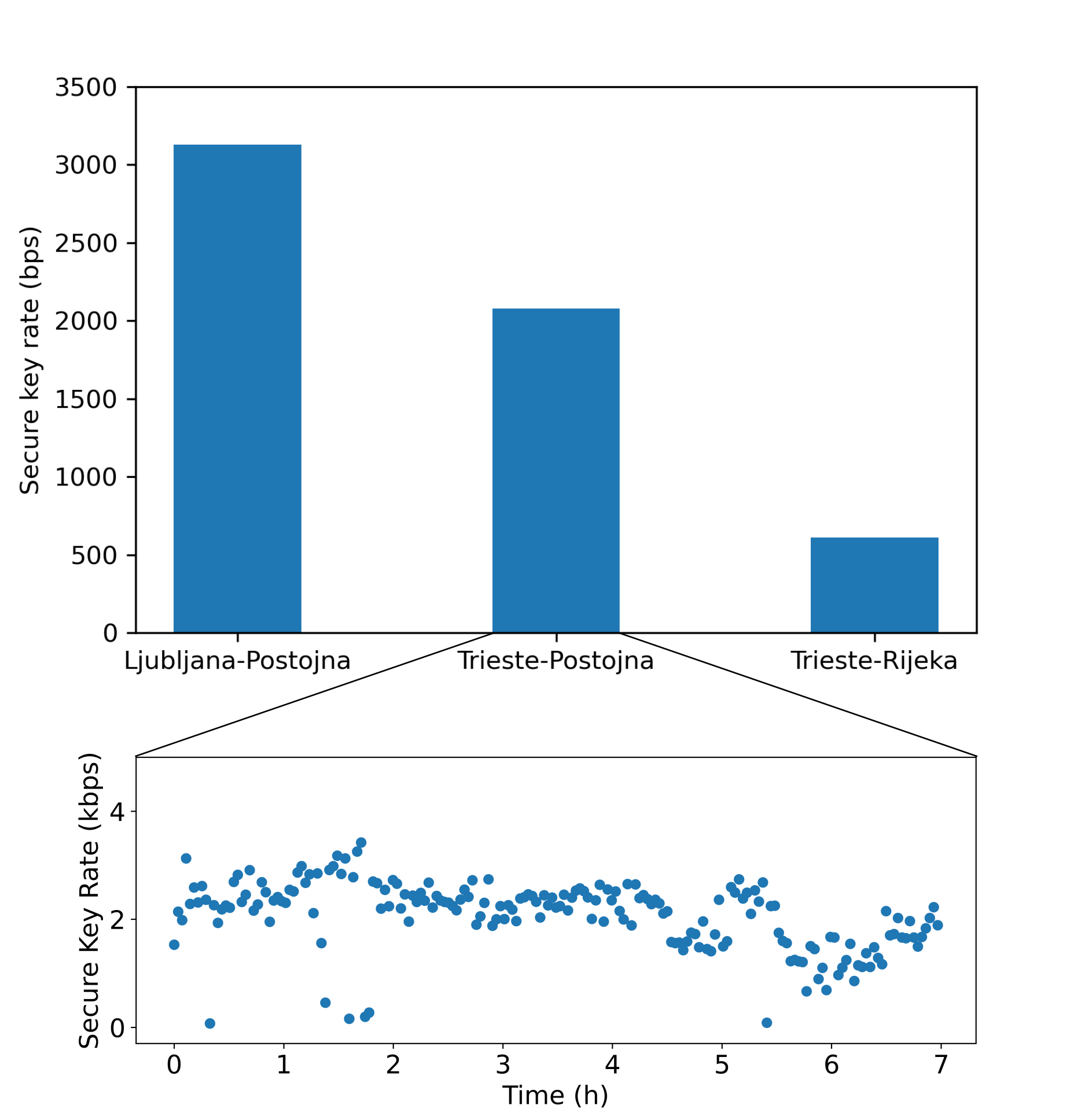}
    \caption{\textbf{Secure key rate and stability of the QKD systems.} The graph on top shows the secret key rate in the three links. \textbf{Inset graph:} the trend of secure key rate in the Trieste-Postojna link for a duration of approximately 7 hours; each point represents 130 seconds of data acquisition, corresponding to the chosen block size.}
    \label{fig:long_meas}
\end{figure}

Figure \ref{fig:long_meas} summarizes the achieved SKRs and long-term stability measurement performed for more than 7 hours. The continuous data acquisition in the Trieste-Postojna channel, which was stabilized with our implemented phase locking system, reflects the performance of the PLL.



\begin{table}[h]
\begin{tabular}{c|c|c|c}
&TS-PO & LJ-PO & TS-RI\\
\midrule
Attenuation (dB) & 14 &  14.3 & 25\\
$\mu_1$ &  0.24 & 0.15 & 0.24\\ 
$\mu_2$ & 0.11 & 0.06 & 0.11\\
Block size $n_Z$ & $1.8\cdot 10^6$ & $1.2\cdot 10^6$ & $6.0\cdot 10^6$\\
Block time (min) & 2.2 & 1.0 & 32.6\\
$\text{QBER}_Z (\%)$ & 1.29 & 0.82 & 2.90\\ 
$\text{QBER}_X (\%)$ & 5.2 & 7.0 & 5.15\\
$SKR (\text{bps})$ & 2080 & 3130 & 610 \\
Temporal Filters (ps)& 100 & 200 & 60\\
$\text{Loss}_Z (\text{dB})$ & 1.4 & 0.2 & 0.8 \\
$\text{Loss}_X (\text{dB})$ & 8.6 & 5.2 & 1.5\\
\end{tabular}
\caption{Specifications of the network and experimental measurements on the three links Trieste-Postojna (TS-PO), Trieste-Ljubljana (TS-LJ) and Trieste-Rijeka (TS-RJ). The data are explained in the text.}
\label{results}
\end{table}



\section{Discussion and Conclusions}
This work represents the first step towards the EuroQCI \cite{euroqci_2}. So far, several countries around the world have already established QKD networks which are up and running and are implemented for various use-cases. e.g., banks, governments, medical centers, etc. In a similar way, Europe is focusing on the development of a European quantum communication network that faces several challenges: multiple vendors, different standards, various implementations of QKD protocols, and ``classical'' infrastructures. This work represents a concrete step towards the realization of such networks, considering that three QKD systems based on the same protocol but realized by different partners (QTI s.r.l., CNR-INO, DTU) have been combined together distributing a quantum key. Despite the fact that the implemented QKD protocol was the same for all the systems, different measurement apparatuses have been adopted in each receiver taking into account the peculiarities of each link. In particular, the Trieste-Rijeka link employed an ultra-low-loss free-space interferometer to partially compensate for the high channel attenuation. Notwithstanding this loss, as high as 25 dB, it was possible to realize a stable QKD link, which is quite difficult to achieve using standard InGaAs detectors. The reason for the lower key rate in the Trieste-Postojna link, with respect to the Ljubljana-Postojna one with similar attenuation (14 dB), is due to the fact that the second link shows a 1.2 dB lower attenuation in the $\mathbf{Z}$-basis, which cannot be fully compensated by the lower $\mathbf{X}$-basis QBER.

It is important to note that the fiber network that we have used has been temporarily established for this specific purpose and this has been possible thanks to the cooperation of the Italian, Slovenian, and Croatian telecommunication operators. In this context, it is crucial to remark that some of the devices we have employed in our demonstration are not optimized for application to   a real telecom-grade QKD system. Possible improvements include a real-time optical switch, which should be preferred for nodes where a transmitter serves more than one user instead of our beam-splitter, and a real-time quantum random number generator (QRNG), which should be used in all Alice's devices.

Finally, we would like to stress two facts. 
First, although a generic network should connect any two users (and not the only adjacent ones), the configuration we have adopted allows to share the paired keys between two parties using a trusted-node strategy. 
For example, Rijeka and Ljubljana, even though not directly connected, can share a key by XORing the keys generated between Trieste-Rijeka, Trieste-Postojna and Postojna-Ljubljana. Second, there is no practical difference between a trusted node made of a switch and a node made of two receiver modules: different nodes produce different keys even if they share the same transmitter since it is the peculiarity of the link (e.g. losses, detector efficiencies, etc.) that generates the exact key.

In conclusion, we have demonstrated an inter-European QKD network during the G20 event held in Trieste. Employing two trusted nodes, we have successfully managed to distribute quantum keys among three different users and to secure a virtual private network among all of them using these keys. This work paves the way towards a fully-fledged European quantum network with concrete use-cases.  


\section{Methods}
To extract a final error-free secret key with guaranteed security, post-processing of the raw time tags registered by Bob is necessary. Post-processing comprises a series of steps explained in detail below.

\subsection{Sifting}

Alice and Bob start the post-processing by sifting the instances of preparation-measurement that match with each other and discarding the rest. At this stage, Bob communicates the time and the basis in which a measurement is registered via the classical physical layer. Note that only the basis chosen by Bob is communicated and not the measurement result in that basis. Alice returns the list of detection events  which Bob measured in the correct basis as well as selected $\mu$ of those instances and they discard the rest. Furthermore, Alice discards instances that are not registered due to loss in the channel. The sifted key, however, may contain errors and correlations with Eve.

The size of the sifted key block for each link has been chosen as the largest one such that the post-processing time is not longer than the acquisition time.

\subsection{Error correction}
A reconciliation process is necessary to guarantee that Alice and Bob shared the same key. We employed a cascade algorithm \cite{bennett1992experimental,brassard1993secret} to remove any errors in the keys that could have been caused by imperfect measurement, noise, or eavesdropper. According to the cascade algorithm, the shared sifted key is split into blocks whose parities are compared and, when a mismatch is found, a dichotomic search is performed in order to identify and correct the mismatching bits. A series of iterations with block sizes progressively doubled has guaranteed the total accordance between the two keys. An initial block size of $k_1\lceil 0.73/QBER \rceil$ has been chosen as proposed in \cite{brassard1993secret}, while a number $n=8$  of iterations has been chosen in order to drastically minimize the failure probability. Although these choices produce a bigger bit leakage than other proposals \cite{vandijk_1997,yan_2008}, they have been preferred in order to minimize the failure probability of this post-processing step \cite{martinez2014demystifying}. A measure of the efficiency of an error correction algorithm is the error reconciliation efficiency $f_{EC}$, defined as \cite{martinez2014demystifying}:
$$f_{EC}=\frac{1-m/n_Z}{H_2(\epsilon)},$$
where $m$ is the length of the message exchanged for the reconciling procedure, $n_Z$ is the block size and $\epsilon$ is the quantum bit error rate. The error reconciliation efficiencies for the three links are reported in table \ref{err_rec_eff}.
\newline
\begin{table}[h]
\begin{tabular}{c|c}
Link & $f_{EC}$\\
\midrule
Trieste-Postojna & 1.28 \\
Trieste-Rijeka & 1.26 \\
Ljubljana-Postojna & 1.25 \\
\end{tabular}
\caption{Error reconciliation efficiency $f_{EC}$ for each link.}
\label{err_rec_eff}
\end{table}

\subsection{Error verification}
The purpose of error-correction is to remove any discrepancies between Alice's and Bob's strings and, hence, it is necessary to verify its success. In order to compare the strings without disclosing it entirely, Alice and Bob calculate the hash value of their error-corrected sifted key and communicates it to each other to check their agreement.



With the desired $\epsilon_{corr}=10^{-12}$, this causes further 40 bits to be discarded in the next privacy amplification step, since $log_2\lceil 1/\epsilon_{corr}\rceil=40$ \cite{canale2014_2}.


\subsection{Privacy amplification}
Upon a positive outcome from error verification step, Alice and Bob share an identical key with error probability $\epsilon_{corr}$. Nevertheless, partial information may be leaked to Eve. \emph{Privacy amplification} (PA) is necessary to minimize Eve's correlation with the sifted and error-corrected key, hence, leaving Alice and Bob with a shared information-theoretic secure key.

PA is achieved by applying a universal hash function in the form of a binary Toeplitz matrix. To form the Toeplitz matrix, Alice generates a string of $n_Z+l-1$ bits and sends it to Bob, where $n_Z$ is the block size and $l$ the final key length, as defined in Eq. (\ref{EQ::}) \cite{kiktenko2016post}. A $l\cross n_Z$ Toeplitz matrix $T$ is built on both sides using the bit string, and finally the dot product $T\cdot key_{sifted}$ produces the privacy-amplified-key. The algorithm has been made less costly in terms of usage of memory and execution time  by avoiding to build the matrix all at once \cite{tang2019high} while adopting parallel executions of the loops.

\subsection{Secure VPN}
The final information-theoretic secure keys, generated by QKD, are ready to be used in crypto devices to encrypt data via standard encryption protocols, such as One Time Pad  (OTP) or Advanced Encryption Standard (AES-256). In this experiment, instead of using commercial crypto devices, the keys were used to establish an end-to-end encryption (E2EE) via a virtual private network (VPN), a technology able to establish a private network over the public internet network. Whit this method, the certificate required by the VPN is securely communicated  between the parties.
An E2EE communication is resilient to eavesdropping as long as the key is undisclosed. 


In this work, a Secure VPN based on OpenVPN software~\cite{openvpn} has been established among all the nodes and its security has been enhanced by QKD by adding a  level of security above and beyond that provided by standard SSL/TLS protocols. Since our VPN used 2048 bit-sized keys, Alice and Bob split their privacy-amplified-keys in blocks of 2048 bits, representing the final keys. 

\section{Acknowledgements}
This work was partially supported by 
the NATO Science for Peace and Security program (Grant No. G5485, project SEQUEL), 
the European Union - PON Ricerca e Innovazione 2014-2020 FESR (Grant no. ARS01\_00734, project QUANCOM), 
the Center of Excellence SPOC - Silicon Photonics for Optical Communications (ref DNRF123), the Innovationsfonden (No. 9090-00031B, FIRE-Q) the EraNET Cofund Initiatives QuantERA within the European Union’s Horizon 2020 research and innovation program (grant agreement No. 731473, project SQUARE), 
the Region Friuli Venezia Giulia (project ``Quantum FVG''),
the H2020 FET Project TEQ (Grant No. 766900), 
the Croatian Science Foundation HRZZ (grant No. IPS- 2020-1-2616) and 
the Croatian Ministry of Science and Education MSE (grant No. KK.01.1.1.01.0001). 
A.R. acknowledges ARRS (grant J2-2514). 
R.K. and \v{Z}.P. acknowledge support by the Slovenian Research Agency (research projects
N1-0180, J2-2514, J1-9145 and P1-0125).
The authors acknowledge, the Italian Ministry for Economic Development (MISE) and the Italian Ministry for Foreign Affairs and International Cooperation (MEACI) for hosting the demonstration during the G20 and for the logistic support, and the Ministry of Foreign and European Affairs of the Republic of Croatia.

The authors acknowledge Tommaso Calarco for his valuable support, TIM S.p.A., Telecom Italia Sparkle S.p.A., Telekom Slovenije d.d., Stelkom and Oda\v{s}ilja\v{c}i i veze d.o.o. (OIV) for making possible all the experiments, providing optical (dark) fiber infrastructure and great technical support. 

The authors acknowledge LightNet, the optical-fiber infrastructure of the research and academic centers of the Region Friuli Venezia Giulia, in particular Prof. Antonio Lanza, for providing the local fiber connections and for the technical support.

 

\bibliographystyle{unsrt}
\bibliography{mybib}

\end{document}